\documentclass{caosp302}

\usepackage{graphicx}

\articleNo{123}
\pubyear{2005}
\volume{35}
\volnumber{3}
\firstpage{1}
\received{May 1, 2005}
\accepted{August 28, 2005}

\begin{document}

\htitle{Magnetism in pre-MS intermediate-mass stars and the fossil field hypothesis}
\hauthor{E.\,Alecian {\it et al.}}

\title{Magnetism in pre-MS intermediate-mass stars and the fossil field hypothesis}

\author{
        E.\,Alecian \inst{1,} \inst{2}
        \and
        G.A.\,Wade \inst{1}
        \and
        C.\,Catala \inst{2}
        \and
        C.\,Folsom \inst{3}
        \and
        J.\,Grunhut \inst{1}
        \and
        J.-F.\,Donati \inst{4}
        \and
        P.\,Petit \inst{4}
        \and
        S.\,Bagnulo \inst{3}
        \and
        S.C.\,Marsden \inst{5}
        \and
        J. Ramirez \inst{2}
        \and
        J.D.\,Landstreet \inst{6}
        \and
        T.\,Boehm \inst{4}
        \and
        J.-C.\,Bouret \inst{7}
        \and
        J.\,Silvester \inst{1}
       }

\institute{
           Royal Military College, CANADA
           \and
           LESIA (Observatoire de Paris), FRANCE
           \and
           Armagh Observatory, U.K.
           \and
           LATT (Observatoire Midi-Pyr\'en\'ees), FRANCE
           \and
           Anglo-Australian Observatory, AUSTRALIA
           \and
           University of Western Ontario, CANADA
           \and
           Laboratoire d'astrophysique de Marseille, FRANCE
          }

\date{March 8, 2003}

\maketitle

\begin{abstract}
Today, one of the greatest challenges concerning the Ap/Bp stars is to understand the origin of their slow rotation and their magnetic fields. The favoured hypothesis for the latter is the fossil field, which implies that the magnetic fields subsist throughout the different evolutionary phases, and in particular during the pre-main sequence phase. The existence of magnetic fields at the pre-main sequence phase is also required to explain the slow rotation of Ap/Bp stars. 
However, until recently, essentially no information was available about the magnetic properties of intermediate-mass pre-main sequence stars, the so-called Herbig Ae/Be stars. The new high-resolution spectropolarimeter ESPaDOnS, installed in 2005 at the Canada-France-Hawaii telescope, provided the capability necessary to perform surveys of the Herbig Ae/Be stars in order to investigate their magnetism and rotation. These investigations have resulted in the detection and/or confirmation of magnetic fields in 8 Herbig Ae/Be stars, ranging in mass from 2 to nearly 15 solar masses. In this contribution I will present the results of our survey, as well as their implications for the origin and evolution of the magnetic fields and rotation.
\end{abstract}

%
\section{Introduction}
\label{intr}

%
\subsection{Problematics}

Among the chemically peculiar (CP) stars, we distinguish the magnetic Ap/Bp stars, hosting large-scale magnetic fields with characteristic strength about 1 kG. Two hypothesis have been proposed to explain the origin of their magnetic fields: the core dynamo hypothesis and the fossil field hypothesis. The core dynamo hypothesis assumes that a magnetic field is generated in the convective core of the main sequence (MS) Ap/Bp stars, and then diffuses towards the surface. On the contrary, the fossil field hypothesis assumes that the stellar magnetic fields are relics from the field present in the parental interstellar cloud or possibly generated in their evolution, by a dynamo that has ceased to operate. It implies that magnetic fields can (at least partially) survive the violent phenomena accompanying the birth of stars, and can also remain throughout their evolution and until at least the end of the MS, without regeneration.

According to the fossil field model, some pre-main sequence (PMS) stars of intermediate mass, the so-called Herbig Ae/Be (HAeBe) stars, should be magnetic. However no magnetic field was observed up to recently in these stars (except in HD~104237, Donati et al., 1997). {\it Can we obtain some observational evidence of the presence of magnetic fields 
in PMS A/B stars
, as predicted by the fossil field hypothesis? If some HAeBe stars are discovered to have magnetic fields, is the fraction of magnetic to non-magnetic HAeBe stars the same as the fraction for main sequence stars? Is the magnetic fields in HAeBe stars strong enough and of appropriate geometry to explain those of the Ap/Bp stars?}

Chemical peculiarities and magnetism are not the only characteristic properties observed in the Ap/Bp stars. Most magnetic MS stars have rotation periods (typically of a few days) that are several times longer than the rotation periods of non-magnetic MS stars (a few hours to one day). It is usually believed that magnetic braking, in particular during PMS evolution when the star can exchange angular momentum with its massive accretion disk, is responsible for this slow rotation (St{\c e}pie{\'n}, 2000). An alternative to this picture involves a rapid dissipation of the magnetic field during the early stages of PMS evolution for the fastest rotators, due to strong turbulence induced by rotational shear developed under the stellar surface, as occurs in solar-type stars (e.g. Ligni\`eres, 1996). In this scenario, only slow rotators could retain their initial magnetic fields, and evolve as magnetic stars to the main sequence. So another question to be addressed is the following: {\it does the magnetic field control the rotation of the star, or does the rotation of the star control the magnetic field?}

To answer to these questions we adopt two strategies. The first consists of observing the brightest "field" HAeBe stars in order to detect magnetic fields, to measure rotation velocity and finally to make a statistical study to be compared to the statistical results of magnetism and rotation of the main sequence A/B stars. The second strategy consists of observing HAeBe stars in young open clusters and associations. Observing stars in a single cluster or association will provide us with stars of the same age and initial conditions (chemical composition, angular mometum, environment). Then, by selecting members of open clusters and associations of various ages, we hope to be able to disentangle "evolutionary" effects from "initial conditions" effects.
Thanks to both studies we will therefore understand the evolution of the magnetic field during the pre-main sequence phase, and its impact on the evolution of the stars.

%
\subsection{The Herbig Ae/Be stars}

The Herbig Ae/Be stars are intermediate-mass pre-main sequence stars, and therefore the evolutionary progenitors of the MS A and B stars. They are distinguished from the classical Be stars by their abnormal extinction law and the association with nebulae, characteristics which are due to their young age. 

They display many observational phenomena often associated with magnetic activity. First, highly ionised lines are observed in the spectra of some stars (e.g. Bouret et. al., 1997), and X-ray emission has been detected coming from the environments of HAeBe stars (e.g. Hamaguchi et al., 2005). In active cool stars, many of these phenomena are produced in hot chromospheres or coronae. Some authors have also mentioned rotational modulation of resonance lines which they speculate may be due to rotation modulation of winds structured by magnetic field (e.g. Catala et al. 1989). 

In the literature we find many clues to the presence of circumstellar disks around these stars, from spectroscopic and photometric data (e.g. Mannings \& Sargent 1997). Recently, using coronographic and interferometric data, some authors have also found direct evidence of circumstellar disks around these stars (e.g. Vink et al. 2002). These disks shows similar properties to the disks of their low mass counterpart (Natta et al. 2001), the T Tauri stars, whose emission lines have been explained by magnetospheric accretion models (Muzerolle et al. 2001). Finally Muzerolle et al. (2004) have sucessfully applied their magnetospheric accretion model to one HAeBe stars to explain the emission lines in its spectrum. 

For all these reasons we suspect that the HAeBe stars may host large-scale magnetic fields that should be detectable with current instrumentation. However, many authors have tried to detect such fields without much success (Catala et al. 1989, 1993; Donati et al. 1997; Hubrig et al. 2004; Wade et al. 2007). The previous instruments were limited either by their spectral range, or by their spectral resolution, or by low efficiency. However, in 2005 a new high-resolution spectropolarimeter, ESPaDOnS, has been installed at the Canada-France-Hawaii telescope. This powerful new instrument has provided us with the capability to sensitively survey many HAeBe stars in order to investigate rotation and magnetism in the pre-main sequence stars of intermediate mass.

%
\section{Observations}

Our data were obtained using the high resolution spectropolarimeter ESPaDOnS installed on the 3.6 m Canada-France-Hawaii Telescope (CFHT, Donati et al., in preparation) during many scientific runs.
ESPaDOnS is 15 to 20 times more efficient than its ancestor MUSICOS, and it covers a large spectral range from 380 to 1080~nm.

We used this instrument in polarimetric mode, measuring Stokes $I$ and $V$ spectra of 65000 resolution. Each exposure was divided in 4 sub-exposures of equal time in order to minimise the impact of instrumental polarisation (Donati et al. 1997; Donati et al. in preparation). The data were reduced using the "Libre ESpRIT" package especially developed for ESPaDOnS, and installed at the CFHT (Donati et al. 1997; Donati et al. in preparation).

We then applied the Least Squares Deconvolution procedure to all spectra (Donati et al. 1997), in order to increase our signal to noise ratio. For each star we used a mask computed using "extract stellar" line lists obtained from the Vienna Atomic Line Database (VALD)\footnote{http://www.astro.univie.ac.at/$\sim$vald/}, with effective temperatures and $\log g$ suitable for each star (Wade et al. in prep.). We excluded from this mask hydrogen Balmer lines, strong resonance lines, lines whose Land\'e factor is unknown and emission lines. The results of this procedure are the mean Stokes $I$ and Stokes $V$ LSD profiles.

%
\section{Field Herbig Ae/Be stars study}

%
\subsection{Our sample in the HR diagram}

\begin{figure}[t!]
\centering
\includegraphics[width=6cm]{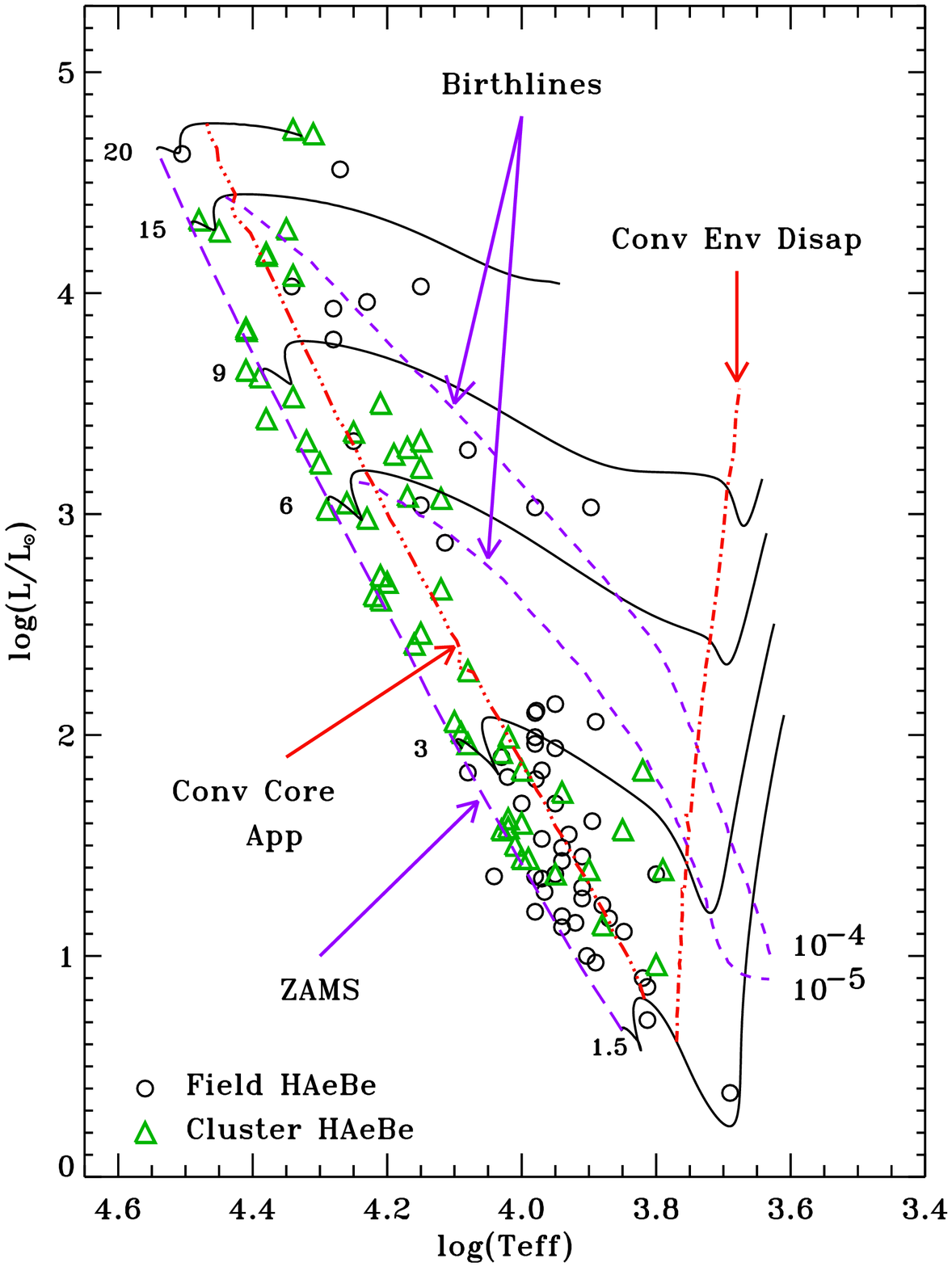}
\hfill
\includegraphics[width=6cm]{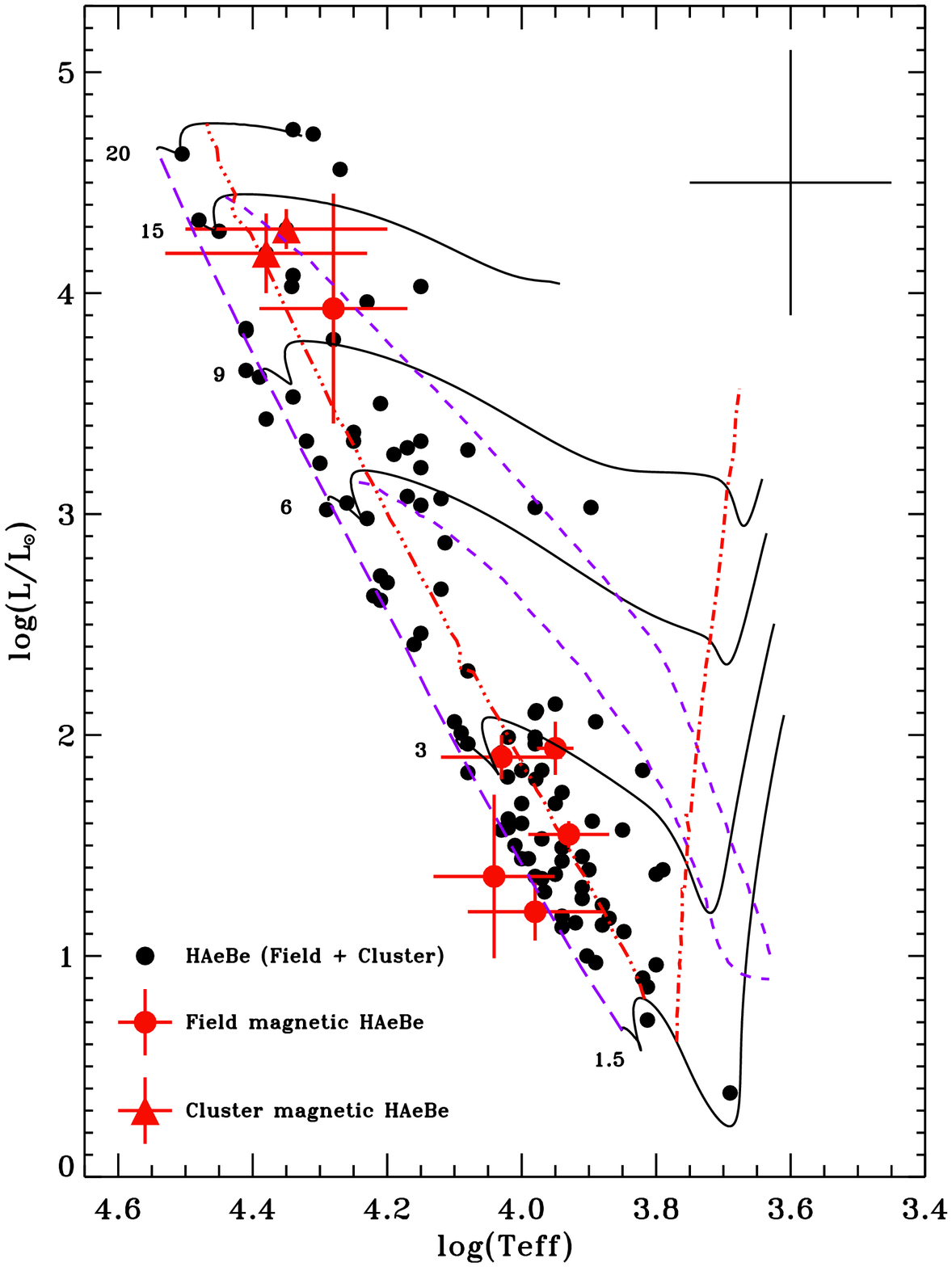}
\caption{{\it Left: } Field HAeBe (circle) and cluster HAeBe (triangle) stars plotted in an HR diagram. {\it Right: } Undetected field and cluster HAeBe stars (small black dots), magnetic field HAeBe stars (big red dots with error bars) and magnetic cluster HAeBe stars (filled triangle with error bars). The large cross in the upper-right of the panel is the mean error bars in luminosity and temperature of the undetected stars. {\it Both panels:} The birthline for 10$^{-5}$ and 10$^{-4}$ $M_{\odot}$.yr$^{-1}$ mass accretion rate are plotted in blue small-dashed line (Palla \& Stahler 1993). The blue long-dashed line is the ZAMS. The PMS evolutionary tracks (full black line), the convective envelope disappearance (red dot-dashed line) and the convective core appearance (red dot-dot-dot-dashed line), calculated with CESAM (Morel 1997) are also plotted.}
\label{fig:hr}
\end{figure}

We have selected HAeBe stars in the catalogues fo Th\'e et al. (1994) and Vieira et al. (1993) with a visual magnitude brighter than 12. Our sample contains 55 stars which have masses ranging from 1.5 M$_{\odot}$ to 15 M$_{\odot}$.
In Fig. \ref{fig:hr} are plotted the stars of our sample in an HR diagram (left panel, circles). For better clarity we plotted in the right panel, a similar HR diagram showing the magnetic HAeBe stars that will be discussed in Sec. 5. The PMS evolutionary tracks and the zero-age main sequence (ZAMS) computed with the CESAM code (Morel 1997), and the birthlines computed by Palla \& Stahler (1993) with two mass accretion rates during the protostellar phase : 10$^{-5}$~M$_{\odot}$.yr$^{-1}$ and 10$^{-4}$~M$_{\odot}$.yr$^{-1}$, are plotted in both panel. The birthline is the locus in the HR diagram of newly formed stars, that is a proto-star becoming optically visible by clearing away its opaque envelope. The PMS lifetime of a star is therefore from the birthline to the ZAMS. 

In Fig. \ref{fig:hr} are also plotted the convective envelope disappearing and the convective core apparition lines, calculated with CESAM. The stars placed between both lines are totally radiative. Our sample contains therefore stars with radiative core and convective envelope, stars with convective core and radiative envelope, and stars which are {\bf totally radiative}. The variety of internal structure of stars in our sample will allow us to test both the fossil field hypothesis and the core dynamo hypothesis.


%
\subsection{The magnetic field}

\begin{figure}[t]
\centering
\includegraphics[width=6.cm, angle=90]{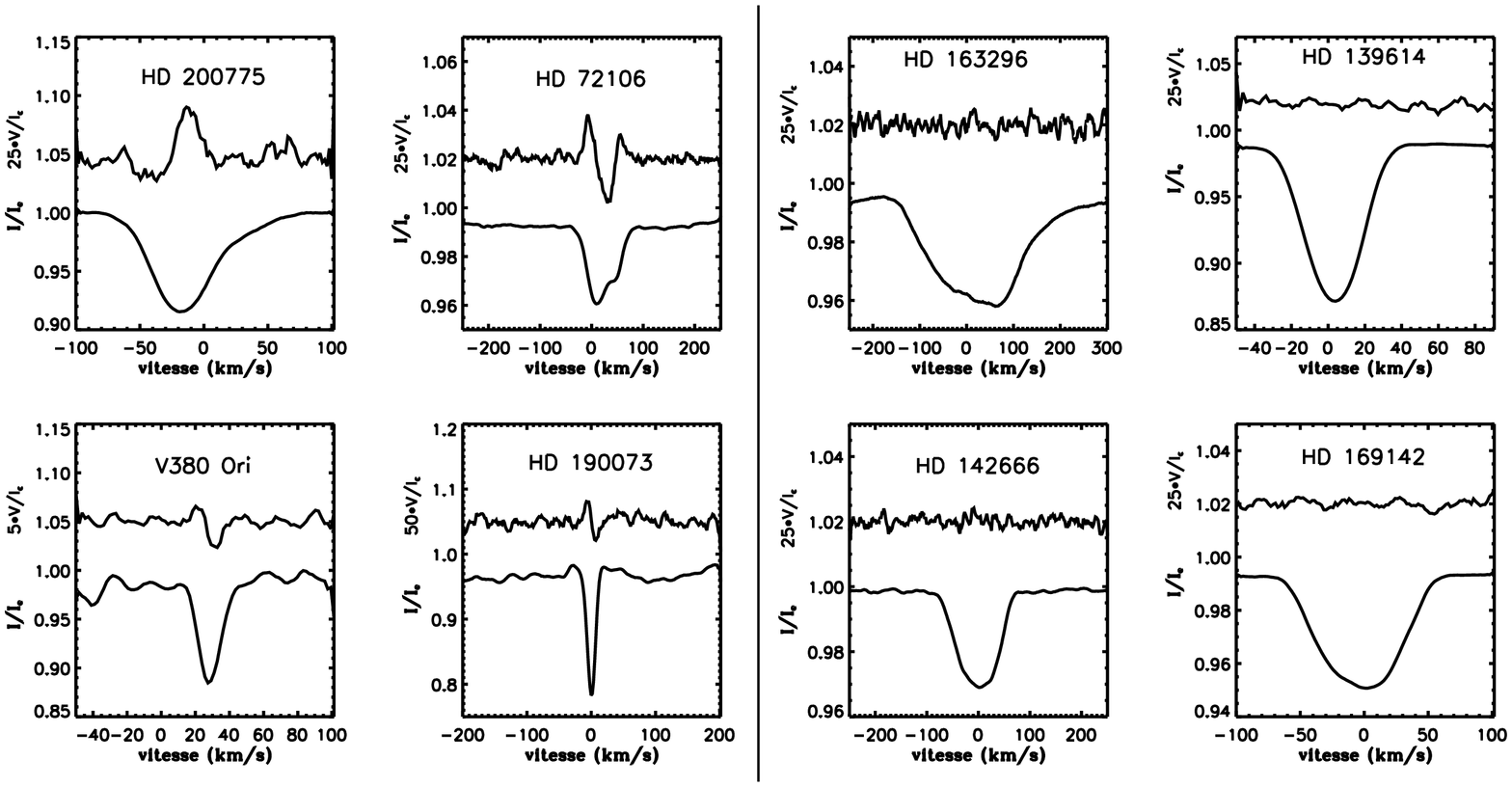}
\caption{Stokes $I$ (bottom) and Stokes $V$ (upper) LSD profiles of the 4 magnetic stars (right) and four undetected stars (left). Note the amplification factor in $V$.}
\label{fig:compil}
\end{figure}

The combination of ESPaDOnS and the LSD method has allowed us to discover four new magnetic HAeBe stars. The Zeeman signatures in the Stokes $V$ profiles of the four magnetic HAeBe stars are plotted in Fig. \ref{fig:compil} on the left. These can be compared to some undetected stars (stars in which we have not detected magnetic fields) on the right. We observe that these signatures are located at the same position as the absorption line, and are as broad as the intensity profile. They are therefore characteristic of the presence of magnetic fields in the stars. As our sample contains 55 stars, the first conclusion that we draw is therefore: ~7\% of our sample HAeBe stars are magnetic.

In order to characterise these magnetic fields, we monitored the four detected stars during many following nights. We recorded variations of the Zeeman signature with time for all stars, except one (HD~190073). To model these variations we assume the oblique rotator model described in Landstreet (1970). This model assumes a dipole of intensity $B_{\rm d}$ placed at a distance $d_{\rm dip}$ of the center of the star, on the magnetic axis inside the star, and inclined with an angle $\beta$ to the rotation axis. The rotation axis make an angle $i$ with the line of sight. As the star rotates with a period $P$, the magnetic configuration at the visible surface of the star, and therefore the Stokes $V$ profile, changes with the rotation phase. We calculated on each point of the observable surface of the star a local intensity $I(\theta,\phi)$ profile using a Gaussian of instrumental width placed at the local velocity. Using the weak field approximation (Landi degl'Innocenti \& Landi degl'Innocenti 1973), we calculated the local Stokes $V(\theta,\phi)$ profile as a function of ${\rm d}I(\theta,\phi)/{\rm dv}$ and the local magnetic field projected on the line of sight $b_{\ell}(\theta,\phi)$. We then integrated $I(\theta,\phi)$ and $V(\theta,\phi)$ over the stellar surface, using the linear limb-darkening law with a coefficient of 0.4 (Claret 2000), in order to get the Stokes $I$ and $V$ profiles. This model is therefore dependent on 6 parameters: the rotation period $P$, the reference time of the rotation phase $T_0$, the magnetic obliquity $\beta$, the inclination of the rotation axis $i$, the dipole intensity $B_{\rm d}$, and the position $d_{\rm dip}$ of the dipole on the magnetic axis with respect to the center of the star.

\begin{figure}[t]
\centering
\includegraphics[width=6.cm, angle=90]{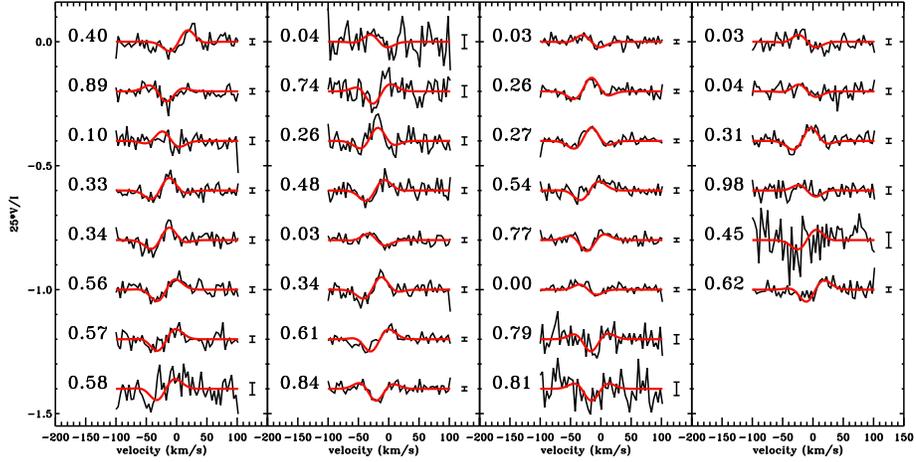}
\caption{Stokes $V$ profiles of HD~200775 superimposed to the synthetic ones, corresponding to our best fit. The rotation phase and the error bars are indicated on the left and on the right of each profile, respectively (Alecian et al. 2007).}
\label{fig:fitv}
\end{figure}

We fitted the observed Stokes $I$ profiles to the synthetic ones, described above, then we calculated a grid of synthetic Stokes $V$ profiles by varying the 6 parameters. Finally, we fitted all observed $V$ profiles of a single star by a $\chi^2$ minimisation. The best model of HD~200775 is superimposed on the observed profiles in Fig. \ref{fig:fitv}, while the best model parameters are summarised in Table \ref{tab:fitv} for all the stars. In the case of V380 Ori we found two possible models. We do not have enough data to choose between them. In all cases the oblique rotator model is sufficient to reproduce our observations.

\begin{table}[t]
\caption{Fundamental, geometrical and magnetic parameters of the magnetic Herbig Ae/Be stars. References : 1 : Alecian et al. (2007), 2 : Folsom et al., in prep., 3 : Alecian et al., in prep., 4 : Catala et al. (2006).}
\label{tab:fitv}
\centering
\begin{minipage}[t]{16cm}
\begin{tabular}{lcccccccc}
\hline
\footnotesize{Star}   & \footnotesize{S.T}. & \footnotesize{P}    & \footnotesize{$B_{\rm P}$} & \footnotesize{$\beta$ ($^{\circ}$)} & \footnotesize{$i$ ($^{\circ}$)} & \footnotesize{$d_{\rm dip}$} & \footnotesize{$B_{\rm P (ZAMS)}$} & \footnotesize{Ref.} \\
         &         & \footnotesize{(d)}  & \footnotesize{(kG)}             &                                 &                         & \footnotesize{$R_*$}             & \footnotesize{(kG)}                          &        \\
\hline
\footnotesize{HD 200775} & \footnotesize{B2}  & \footnotesize{4.3281}         & \footnotesize{1}             & \footnotesize{55}            & \footnotesize{60}            & \footnotesize{0.05} & \footnotesize{3.6}                                             & 1 \\
\footnotesize{HD 72106}   & \footnotesize{A0}  & \footnotesize{0.63995}     & \footnotesize{1.3}          & \footnotesize{60}            & \footnotesize{23}            & \footnotesize{0}    & \footnotesize{1.3}                                             & \footnotesize{2} \\
\footnotesize{V380 ori} 
& \footnotesize{A2}  & \footnotesize{$[7.6,9.8]$} & \footnotesize{1.4}          & \footnotesize{$[90,85]$} & \footnotesize{$[36,49]$} & \footnotesize{0}    & \footnotesize{2.4} & \footnotesize{3} \\
\footnotesize{HD 190073}
& \footnotesize{A2}  &                   & \footnotesize{$[0.1,1]$} & \footnotesize{$[0,90]$}   & \footnotesize{$[0,90]$}   &       & \footnotesize{$[0.4,4]$}                                    & \footnotesize{4} \\
\hline
\end{tabular}
\end{minipage}
\end{table}

This method is very efficient if we observe variations in the Stokes $V$ profiles. However in the case of HD~190073, no such variations are observed. To explain this we propose 3 hypotheses:  either the star is seen pole-on, or the magnetic obliquity is null and the magnetic field is axisymetric, or the rotation period of the star is very long. All these hypothesis are consistent with a dipolar field inside the star, which is further strengthened by the fact that the Zeeman signature is a simple dipolar signature which is stable over more than 2 years.

\begin{figure}[t]
\centering
\includegraphics[width=3.3cm, angle=90]{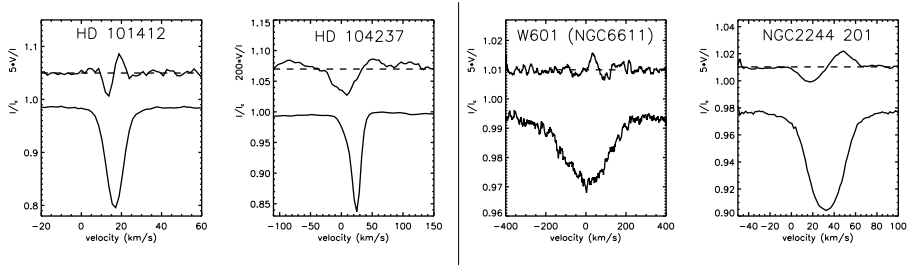}
\caption{Stokes $I$ (bottom) and Stokes $V$ (up) LSD profiles plotted for HD 104237, HD 101412, W601 and NGC~2244~201 (Wade et al. in preparation; Alecian et al. in preparation).}
\label{fig:autre}
\end{figure}

Finally, we mention two other detections obtained using the SemelPol polarimeter and the UCLES spectrograph at the Anglo-Australian Telescope (AAT) (Semel et al. 1993, Donati et al. 1997). The first is HD~104237, which was discovered to be magnetic by Donati et al. (1997), and which has been recently confirmed (Wade et al., in preparation). The second is HD 101412, discovered to be magnetic by Wade et al. (2007) using FORS1 at VLT, and which has been recently confirmed using SemelPol. The Zeeman signatures are plotted in Fig. \ref{fig:autre}, on the left, and are typical signatures of a magnetic field inside the star. We still don't have enough data to characterise their magnetic fields, but the simple signatures are consistent with organised fields.

%

%
\section{Herbig stars in young open clusters}

Motivated by all these results we decided to go further and to observe HAeBe stars in young open clusters and associations. We selected Herbig members in three clusters: NGC 2244, NGC 2264 and NGC 6611 from the catalogues of Park \& Sung (2002), Park et al. (2000) and de Winter et al. (1997).

We observed 16 stars in NGC 2244, 25 stars in NGC 2264 and 17 stars in NGC 6611, which are plotted  in Fig. 1 (left). We detected magnetic fields in stars of two different clusters: NGC~2244~201 and W601 (NGC~6611) (Fig. \ref{fig:autre} right), but none in NGC 2264. To explain the lack of detection of magnetic fields in the latter, we propose that the initial conditions of star formation may play a role in the intensity of the magnetic fields of A/B stars.

The detection in W601 is very interesting, as this is our only detection of magnetic field in a fast rotator (${\rm v}\sin i\sim180$~km.s$^{-1}$). Furthermore, this star is very young ($<1\;$Myr) compared to NGC~2244~201 ($\sim6\;$Myr). Both stars have similar spectral type: B1.5 and B1 respectively, but W601 rotates with  ${\rm v}\sin i\sim180$~km.s$^{-1}$, while NGC~2244~201 rotates with only ${\rm v}\sin i\sim25$~km.s$^{-1}$ (Alecian et al. in preparation). These stars may provide us with clues about angular momentum evolution during the pre-main sequence phase.

%
\section{Conclusion}

In order to investigate magnetism in the PMS stars of intermediate mass, we performed a survey of 55 field HAeBe stars using the new high-resolution spectropolarimeter ESPaDOnS (CFHT). These investigations result in the detection of 4 magnetic fields, leading to our first statistical conclusion:  approximately 7\% of HAeBe stars are magnetic. According to the distribution of magnetic Ap/Bp stars among the MS A/B star, and assuming a fossil field hypothesis, we expect between 2 and 10 \% of HAeBe stars (Power et al. in preparation) to be magnetic, consistent with our result.

We have monitored the 4 magnetic stars over many nights in order to determine the strength and topology of their magnetic fields. We conclude that for 3 of them (HD 200775, HD 72106, V380 Ori) the field can be described by a dipole placed at the center of the star or slightly decentered inside the star, with strength ranging from 1~kG to 1.4~kG. In the case of HD~190073 we are not able to characterise its magnetic field as the Stokes $V$ profile does not vary. However, we conclude that the simple dipolar signature of the Stokes $V$ profile, which is stable over more than 2 years, is consistent with a large-scale organised magnetic field, with strength between 100 G and 1 kG.

At the same time, we observed two other southern-hemisphere HAeBe stars using the UCLES spectrograph and the SemelPol polarimeter. We confirm the detections in HD~104237 and in HD~101412 announced by Donati et al. (1997) and Wade et al. (2007). Although we do not have enough data to characterise their magnetic fields, their simple dipolar signatures are consistent with large-scale organised magnetic fields, similar to those observed in the Ap/Bp stars.

As magnetic HAeBe stars evolve toward the MS, their radii decrease, and their surface magnetic field strengths should increase according to the magnetic flux conservation hypothesis. The projection of the magnetic intensities of the 4 magnetic HAeBe on the main sequence predicts MS magnetic strengths ranging from 400 G to 4 kG (Table \ref{tab:fitv}), which are close to the magnetic intensities of typical Ap/Bp stars. All these results therefore bring strong arguments in favour of the fossil field hypothesis.

We then started to perform a survey of HAeBe stars in 3 young open clusters: NGC~2244, NGC~2264, and NGC~6611. We have detected two magnetic stars in two different clusters (NGC~2244 201 and W601 in NGC~6611), but none in NGC~2264. We may have a clue of the role of the initial conditions of star formation in the intensity of the magnetic fields of A/B stars. In addition, we observe that W601 and NGC~2244 201 are both of similar spectral type (B1.5 and B1), that the first one (1 Myr) is younger than the other (6 Myr), and that W601 is a fast rotator (180 km/s) compared to NGC~2244 201 (25 km/s). We may see a sign of the evolution of angular momentum during the pre-main sequence phase of intermediate mass stars.

Finally in order to confront the fossil field hypothesis to the core dynamo hypothesis, we plotted in the HR diagram of Fig. \ref{fig:hr} (right panel) all the magnetic HAeBe stars, as well as the undetected field and cluster stars. We observe that 5 stars among the 8 magnetic HAeBe stars are in the {\bf totally radiative zone}. Although the error bars are large, one of these is confidently in the totally radiative zone, including its error bars on temperature and luminosity. On the other hand, all 8 magnetic HAeBe stars host the same type of magnetic fields: a large-scale organised magnetic field. We therefore conclude that presence of core convection does not appear to be responsible for the presence of magnetic fields in HAeBe stars. The magnetic fields of the intermediate mass stars are therefore {\bf very likely of fossil} origin.

%
\acknowledgements

EA is founded by the Marie-Curie FP6 program. GAW and JDL are supported by NSERC and DND-ARP (Canada).

\end{document}